# TITLE

**AI and analytics in sports: Leveraging BERTopic to map the past and chart the future**

## AUTHOR'S NAME

Manit Mishra

ORCID: https://orcid.org/0000-0002-2813-0064

## SHORT BIOGRAPHY

Dr. Manit Mishra is Professor (Marketing) and Dean (Research & Accreditation) at IMI Bhubaneswar, Odisha, India. He is recipient of the prestigious AIMS-ICFAI Best Teacher in India award for the year 2023. He teaches Marketing Analytics, Marketing Research, and Consumer Behavior. His areas of research interest are marketing modelling; and applying statistical as well as machine learning methods to structured and unstructured marketing data. His research profile includes nine ABDC 'A' category publications. He is Associate Editor of the journal *Global Business Review*.

## CONTACT DETAILS OF THE AUTHOR

Dr. Manit Mishra
Professor (Marketing) and Dean (Research & Accreditation)
IMI Bhubaneswar
IDCO Plot # 1, Gothapatna, PO: Malipada
Bhubaneswar – 751003
Odisha, India
Ph: +91 8658988485
E-mail: manit.mishra@imibh.edu.in

# AI and analytics in sports: Leveraging BERTopic to map the past and chart the future

## Abstract

**Purpose:** The purpose of this study is to map the body of scholarly literature at the intersection of artificial intelligence (AI), analytics and sports and thereafter, leverage the insights generated to chart guideposts for future research.

**Design/methodology/approach:** The study carries out systematic literature review (SLR). Preferred Reporting Items for Systematic Reviews and Meta-Analysis (PRISMA) protocol is leveraged to identify 204 journal articles pertaining to utilization of AI and analytics in sports published during 2002 to 2024. We follow it up with extraction of the latent topics from sampled articles by leveraging the topic modelling technique of BERTopic.

**Findings:** The study identifies the following as predominant areas of extant research on usage of AI and analytics in sports: performance modelling, physical and mental health, social media sentiment analysis, and tactical tracking. Each extracted topic is further examined in terms of its relative prominence, representative studies, and key term associations. Drawing on these insights, the study delineates promising avenues for future inquiry.

**Research limitations/implications:** The study offers insights to academicians and sports administrators on transformational impact of AI and analytics in sports.

**Originality/value:** The study introduces BERTopic as a novel approach for extracting latent structures in sports research, thereby advancing both scholarly understanding and the methodological toolkit of the field.

**Introduction**

The transformative power of analytics in sports first captured mainstream attention with the release of the 2011 film Moneyball, starring Brad Pitt (Elitzur, 2020; Swartz, 2020). The movie, based on Michael Lewis's 2003 book of the same name, narrates the pioneering journey of Billy Beane, General Manager of the Oakland Athletics baseball team, and his adoption of Sabermetrics, a form of sports analytics, to build a competitive team despite budgetary constraints. Since then, analytics has not only evolved into an indispensable tool but has also served as a harbinger for leveraging artificial intelligence (AI) in sports. Sports presents a fertile ground for such applications, owing to its rich repositories of data and the ever-increasing need for strategic decision-making (Sarlis & Tjortjis, 2020; Swartz, 2020). The impact of AI and analytics is multifaceted and extends across several domains within the sporting ecosystem. These include strategic decision-making (Elitzur, 2020), performance optimization of players (Casals & Daunis-i-Estadella, 2023), talent identification and recruitment (Maanijou & Mirroshandel, 2019), fan experience management (Liu *et al.*, 2022), and injury prevention and recovery (Dorschky *et al.*, 2023).

Hulland and Houston (2020, p. 351) assert that "science advances by building on prior knowledge," underscoring the importance of synthesizing existing research to inform future inquiry. However, despite the growing scholarly interest in the application of AI and analytics within the sports domain, a comprehensive and methodologically robust synthesis of the field remains conspicuously absent. Such a review would facilitate an improved understanding of the current scope, thematic structure, and avenues for future research. Existing review studies exhibit notable limitations in terms of at least three aspects that constrain the advancement of academic understanding as well as practical implementation. First, current literature reviews tend to adopt narrow topical or disciplinary scopes. For example, Ghosh *et al.* (2023) focus exclusively on sports analytics applications within sensor technologies, computing, and mobile systems, thereby overlooking other related areas within the discipline of sports. Similarly, Goes *et al.* (2021) restrict their review to research utilizing positional tracking data to study tactical behavior in soccer, and Rajšp and Fister Jr (2020) provide a review of smart sport training. While these studies offer valuable insights, their specialized focus limits a holistic understanding of AI and analytics across diverse sporting contexts, thereby constraining both the breadth of insights and the identification of actionable future research pathways. Second, many of existing review papers lack adherence to structured review protocols, diminishing the transparency and reproducibility of their findings (e.g., Chmait & Westerbeek, 2021; Ghosh *et al.*, 2023; Swartz, 2020). Third, and most critically, no prior reviews have employed advanced computational techniques such as topic modeling to synthesize this fast-expanding body of knowledge. For example, studies by Dindorf *et al.* (2023) and Zhao *et al.* (2022) restrict the scope of review to bibliometric analysis. As Snyder (2019) argues, review methodologies must evolve alongside the disciplines they examine, especially in data-intensive fields like AI, analytics, and sports. Topic modelling, a machine learning-based text mining method, allows for the extraction of latent thematic structures from large corpora of scholarly literature. Its application enhances objectivity, uncovers hidden research clusters, and provides a more nuanced understanding of thematic trends. Recent scholars (e.g., Asmussen & Møller, 2019; Mishra, 2025) have emphasized the value of topic modeling in literature reviews, particularly for mapping conceptual landscapes and identifying emergent research frontiers. Given these gaps, there is a compelling need for a systematic, comprehensive, and computationally enriched review of AI and analytics in sports. Such a study not only contributes to theory-building and scholarly coherence but also equips practitioners with evidence-based insights into key areas such as performance modelling, tactical analysis, health and injury prevention, and fan engagement. Moreover, this is the opportune moment for any exercise towards consolidation of knowledge pertaining to research on AI and analytics in sports due to its rapidly burgeoning utilization (Chmait & Westerbeek, 2021; Dindorf *et al.*, 2023; Elitzur, 2020; Swartz, 2020; Zhao *et al.*, 2022).

To address the aforementioned gaps and advance the scholarly discourse on application of AI and analytics in sports, this study adopts a dual-method approach that combines methodological rigor with computational innovation. First, we conduct a systematic literature review (SLR) in accordance with the Preferred Reporting Items for Systematic Reviews and Meta-Analyses (PRISMA) framework (Moher *et al.*, 2009), ensuring transparency, reproducibility, and methodological integrity. This protocol guides the selection of peer-reviewed journal articles published between 2002 and 2024 that focus on AI and analytics driven research within the sports domain. Second, to uncover and analyze the thematic structure of this research landscape, we apply BERTopic - a state-of-the-art topic modelling algorithm based on transformer embeddings and class-based TF-IDF (Grootendorst, 2022). This approach enables the extraction of coherent and interpretable topic clusters from the corpus, facilitating a granular understanding of how the field has evolved over the past two decades. In doing so, the study responds directly to the call by Klaus and Kuppelwieser (2021) towards leveraging of emerging algorithmic techniques for research. Building upon this methodological foundation, the study pursues two primary objectives: (a) to develop a comprehensive and integrative synthesis of existing research on the use of AI and analytics in sports, and (b) to leverage the identified thematic insights to propose strategic guideposts for future research, addressing both conceptual gaps and emerging opportunities in this rapidly evolving field.

This study offers two distinct contributions to the literature. First, it advances the existing body of knowledge on the application of AI and analytics in the sports domain by employing a dual-method approach that integrates a systematic literature review (SLR) with topic modelling. This multi-method design enables a rigorous, structured, and holistic synthesis of scholarly work, aligning with calls for greater methodological robustness in literature reviews (Mishra & Mund, 2024; Vanhala et al., 2020). Second, the study introduces BERTopic, an advanced topic modeling technique grounded in transformer-based embeddings and class-based TF-IDF (Grootendorst, 2022), as a methodological innovation for researchers in sports analytics. By doing so, it not only enhances the analytical depth of this review paper but also encourages the adoption of cutting-edge natural language processing tools in future research within the field. Thus, this study provides a novel input to the domain, a key condition for selection of review topic (Paul and Criado, 2020). In Figure 1, we present the complete workflow utilized towards execution of this paper.

**Figure 1.** Research workflow

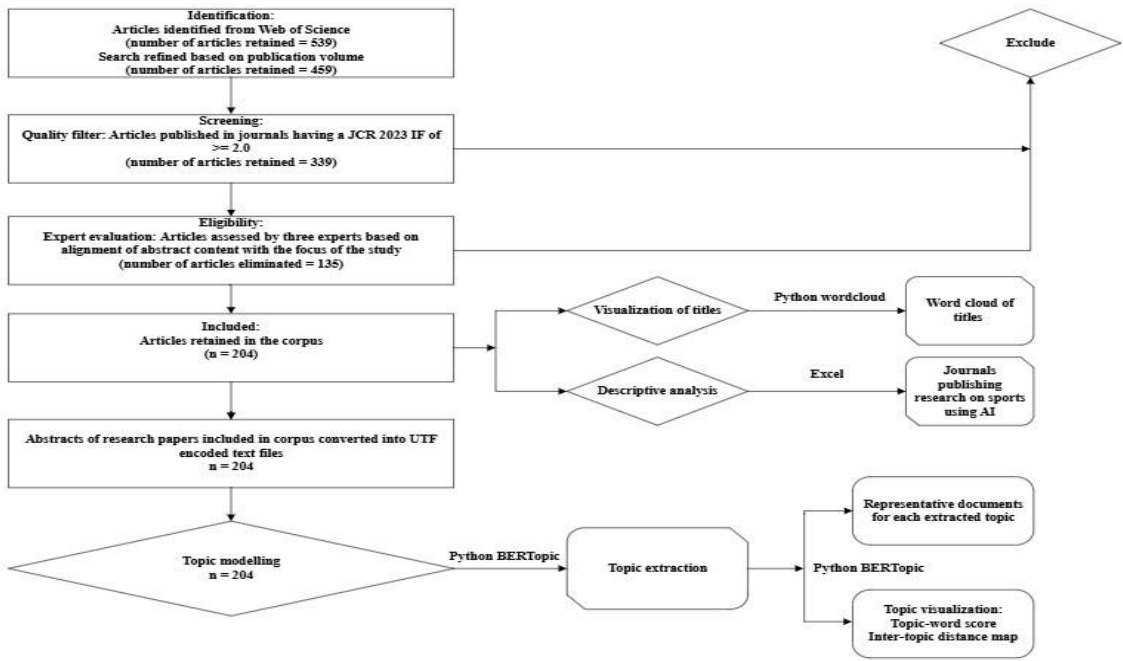

Source: Author's own creation

**Conceptual overview**

"In God we trust. All others must bring data." This quote attributed to W. Edward Demings lies at the genesis of analytics and AI. The growing availability of quality data from varied sources, and access to algorithms that can mine it for deriving insights have facilitated its popularity. Analytics refers to "use of data, statistical and quantitative analysis, explanatory and predictive models, and fact-based management to drive decisions and actions" (Davenport & Harris, 2007). Experts have recognized the relevance of analytics towards improving performance in myriad functional domains such as, marketing (Germann *et al.*, 2013), retail (Bradlow *et al.*, 2017), consumer behaviour (Erevelles *et al.*, 2016), supply chain (Seyedan & Mafakheri, 2020), and finance (Broby, 2022). To cope with the ever increasing requirement for using analytics to derive insights from large and complex real-time data, gradually AI-powered analytics is being leveraged (Basole *et al.*, 2024). Stryker and Kavlakoglu (2025) define AI as the "technology that enables computers and machines to simulate human learning, comprehension, problem solving, decision making, creativity and autonomy." AI, encompassing methods such as machine learning, deep learning, and natural language processing, enables systems to autonomously detect patterns, generate predictions, and continuously learn from high-volume, high-velocity data. Application of AI-powered analytics in areas such as healthcare (Bag *et al.*, 2023), agriculture (Spanaki *et al.*, 2021), education (Chen *et al.*, 2022), and business (Sjödin *et al.*, 2023). As analytics and AI-powered analytics continue to gain traction, decision-makers in sports are also increasingly using it to derive insights (Herold *et al.*, 2024; Qian *et al.*, 2024; Watanabe *et al.*, 2021).

**Methodology**

*Article selection and corpus*

The study extracts the topical structure of extant research on usage of AI in sports. The PRISMA protocol (Moher *et al.*, 2009) and guidance from established studies (Hartmann *et al.*, 2024; Palmatier *et al.*, 2018; Stead *et al.*, 2022) informed our article search and selection process. At the *Identification* stage, consistent with prior studies (Mishra, 2025; Vanhala *et al.*, 2020), we used the Web of Science (WoS) platform to select relevant journal articles, and employed a comprehensive keyword based search (Hiebl, 2021). WoS's adherence to Garfield's Law of Document Sets and Bradford's Law (Borgman & Furner, 2002) ensures high-quality literature curation (Jacsó, 2005), making it a robust platform for article selection. We carried out a search for relevant articles on WoS Core Collection – Social Science Citation Index (SSCI) published during 2002-2024 using the keywords "Artificial Intelligence" OR "Analytics" OR "Machine Learning" OR "Deep Learning" OR "AI" OR "ML" AND "Sports" within the "*Topic*" field. These keywords were selected to ensure a broad technological scope, capturing both formal and abbreviated terminology commonly used in scholarly discourse on the application of artificial intelligence in sports. The search resulted in identification of 598 articles published during January 2002 to December 2024. We limited our search to journal articles which further reduced the relevant output to 539. To enhance focus and relevance, we refined our search by research area, restricting results to the ten disciplines with the highest publication volume in the domain of AI and analytics in sports. This yielded 459 articles for analysis. At the *Screening* stage of PRISMA protocol, we implemented a journal quality filter to ensure inclusion of high-quality, peer-reviewed research (Hiebl, 2021; Paul *et al*., 2021; Raddats *et al*., 2019). Drawing upon Stead *et al.* (2022), only articles published in journals with a 5-year Journal Citation Reports (JCR) - 2025 Impact Factor (IF) of 2.0 or higher were retained for analysis. This process yielded 339 articles. Thereafter, at the *Eligibility* stage, following Hartmann *et al*. (2024) three academicians read each article abstract and assigned it a binary coding for focus on AI and analytics in sports (1: focused, 0: no clear focus). Any discrepancy in opinion was sorted out by reading the full paper followed by discussion (Tranfield *et al*., 2003). Such a rigorous assessment of eligibility led to exclusion of 135 articles. These studies were excluded because their focus did not involve AI or analytics based algorithms; instead, they primarily employed traditional statistical methods or experimental research designs to address their research objectives. Finally, at the *Included* stage the remaining 204 articles published during 2002-24 in 101 different journals were sampled as part of the corpus for topic modelling.

***Descriptives***

To assess the sample's reflection of the core aspects of this study – "artificial intelligence," "analytics," and "sports," we created a word cloud visualization of the titles of 204 articles in the corpus. The word cloud visualization was formed using using Python wordcloud 1.8.2.2 (Oesper *et al*., 2011) package. In a word cloud, the size of each word corresponds to its frequency or relative prominence within the analysed text corpus. The visualization puts forth "sport," "model," "data," "player," "machine learning," and "performance," as some of the significant words in the corpus (Figure 2). The word cloud output suggests that the sample of 204 papers represents research within the sports domain focused on the application of AI and analytics.

The significance of AI and analytics in sports is evident from the wide range of journals that have published such research (Figure 3). The scope of these journals cover areas such as psychology, sports science, forecasting, analytics, and sustainability. The largest number of articles appear in the journal titled *Frontiers in Psychology* which published 19 research papers during the period 2002-24. The *International Journal of Sports Science & Coaching*, a sports specific journal, published the next highest number of articles (15) followed by *International Journal of Forecasting* (6). The journals *Sports Management Review*, *Big Data*, *PLOS One*, *Applied Sciences-Basel*, and *Sustainability* published the next highest number of 5 articles each. Publications in such acclaimed journals underscore relevance of AI and analytics in sports.

**Figure 2.** Word cloud of the titles of in the corpus

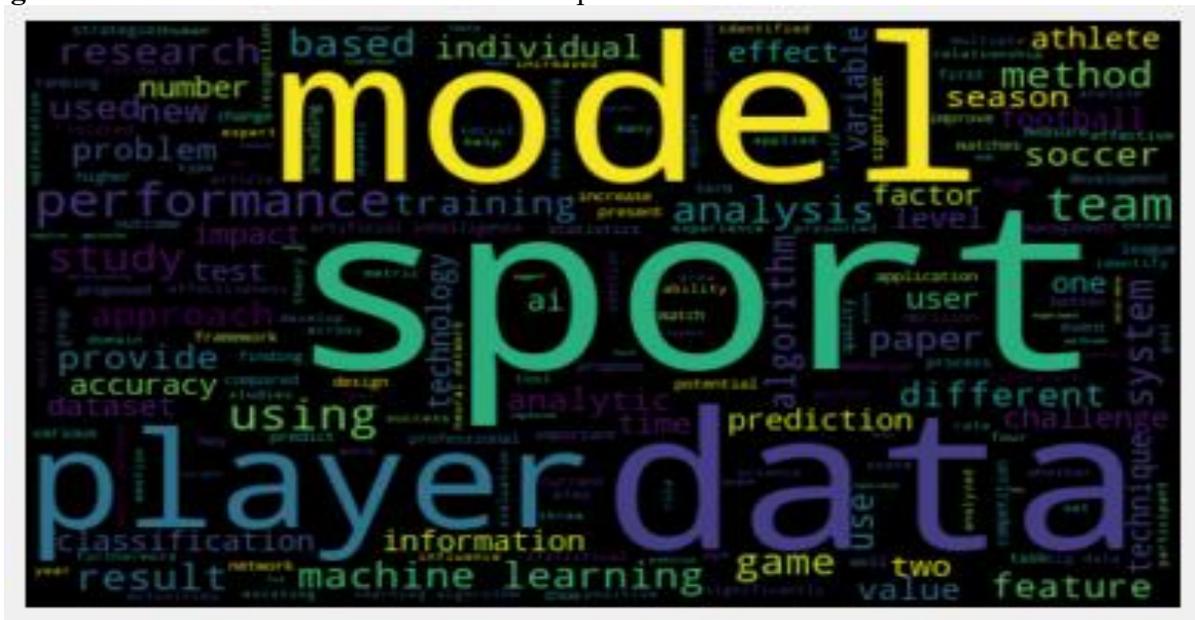

Source: Author's own creation

**Figure 3.** Top journals in terms of output on AI-based research on sports

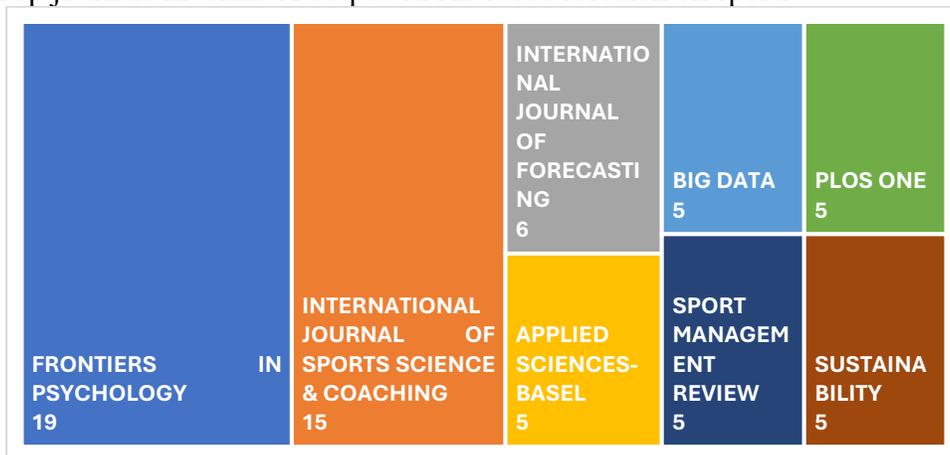

Source: Author's own creation

*Topic modelling*

We carried out topic modelling of extant research on usage of AI and analytics in sports so as to produce visualization of its topical space. Jacobi *et al*. (2016, p. 90) define topic models as "computer algorithms that identify latent patterns of word occurrence using the distribution of words in a collection of documents." Topic modelling employs natural language processing and machine learning techniques to uncover latent semantic patterns within large collections of textual data (Blei *et al*., 2003). We extracted the embedded topics using the unsupervised topic modelling technique of BERTopic (Grootendorst, 2022). BERTopic "leverages clustering techniques and a class-based variation of TF-IDF to generate coherent topic representations" (Grootendorst, 2022, p. 2). BERTopic begins by generating contextualized text embeddings using a pre-trained BERT model, which captures semantic relationships within the data. These embeddings are then clustered to identify coherent topic structures. We selected BERTopic for topic modelling due to its demonstrated effectiveness in producing contextually rich and semantically coherent topics from textual corpora (Grootendorst, 2022). The study is in alignment with prior research to map a domain using topic modelling, in general (e.g., Mishra, 2022; Mishra & Mund, 2024; Tirunillai & Tellis, 2014), and the advanced algorithm of BERTopic, in particular (Mishra, 2025; Mishra & Mohapatra, 2025). For executing topic modelling using BERTopic, the 204 abstracts in the corpus were transformed into UTF-8 encoded text tiles and thereafter, subjected to preprocessing. As recommended by Grootendorst (2022), CountVectorizer was employed to preprocess the documents following the generation of embeddings and clustering. This step is essential as BERTopic, being a transformer-based model, requires complete textual input to produce accurate contextual embeddings. The refined corpus was subsequently analysed using BERTopic resulting in extraction of four topics along with associated terms in decreasing order of relevance (Figure 4), representative articles for each extracted topic (Table 1), and inter topic distance map reflecting all the extracted topics (Figure 5).

*Topic interpretation*

DiMaggio *et al*. (2013, p. 582) opine that the objective while topic modelling extant research in a domain, should not be "to estimate population parameters correctly, but to identify the lens through which one can see the data most clearly." We have endeavoured to contribute a topic model representing extant research such that the focus is on interpretability and analytic utility (Blei & Lafferty, 2009; DiMaggio *et al*., 2013). In alignment with extant research (e.g., Mishra, 2022; Mishra & Mund, 2024), interpretation of topics was done through a thorough assessment of topic-word associations (Crain *et al*., 2012; DiMaggio *et al*., 2013; Hannigan *et al*., 2019) given in Figure 4. The objective of this exercise was to look for the logical connection between the representative words of each topic (Guo *et al*., 2017), and thereafter, label it so as to give more importance to the words having higher weightage (Mishra, 2025). The representative studies for each topic (Table 1) was also assessed to obtain clarity regarding representativeness of the topic labels. For example, the words associated with topic 1 (Figure 4) include "performance," "team", "football," "results," "match," and "player." The representative studies are primarily focused on applying machine learning techniques to analyze and predict player and team performance, particularly in soccer (football). Therefore, the topic was labelled as "performance modelling."

Through a similar process, the rest of the topics were labelled as "physical and mental health" (topic 2), "social media sentiment analysis" (topic 3), and "tactical tracking" (topic 4). The topic physical and mental health encompasses research exploring AI and analytics driven interventions in athlete well-being, injury prevention, and sports education, particularly within collegiate and training environments. Similarly, the third topic of social media sentiment analysis encompasses studies that apply AI and analytical techniques to assess emotional tone, fan reactions, and engagement patterns on platforms such as Twitter, providing insights into digital fan experience and public discourse in sports. And the final topic of tactical tracking encompasses studies that apply AI and analytics to analyze movement data, tactical behaviour, and in-game decision-making, particularly in soccer and other ball sports.

**Figure 4.** Visualization of topics and relevant terms

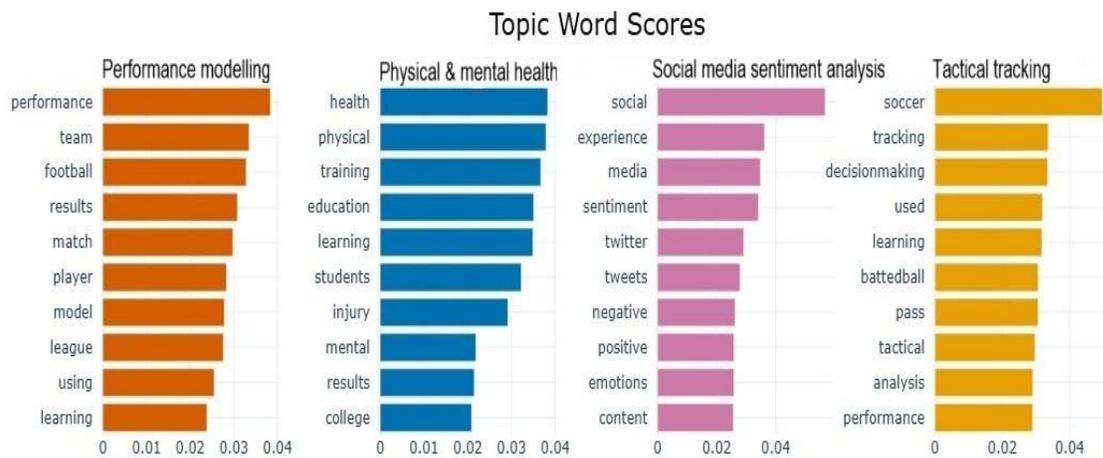

Source: Author's own creation

*Representative studies*

An examination of studies representing each extracted topic further serves towards understanding and validating the obtained topical solution. Topic 1 relates to performance modelling of players using sports data. Maanijou and Mirroshandel (2019) is a representative study which employs a weighted-voting ensemble algorithm, optimized through a genetic algorithm, to predict the ranking of soccer players in the Persian Gulf Premier League. This work demonstrates how AI-driven ensemble methods can surpass traditional statistical approaches, offering more robust tools for player evaluation, ranking, and talent identification. Another representative study on performance modelling is by Al-Asadi and Tasdemir (2022), who demonstrate that machine learning algorithms can provide an objective and quantitative approach to estimating soccer players' market value. Given the central role of market value in transfer negotiations between clubs and player agents, their model offers a valuable baseline for supporting evidence-based decision-making in the negotiation process. The third representative study on performance modelling is by Constantinou (2019), which develops a machine learning model to predict soccer match outcomes in one country by leveraging match results from multiple other countries. This cross-national approach highlights the potential of AI techniques to generalize performance insights across different competitive contexts. The representative studies collectively highlight the role of AI and analytics driven approaches to generate performance insights across different competitive contexts.

Topic 2 encompasses research addressing issues related to physical and mental health. A representative study is Peng and Tang (2021), who employ AI and big data analytics to design a Smart Sports Classroom (SSC) aimed at fostering engagement in physical activity. Their work underscores the role of technology-enhanced sports education in promoting healthy exercise habits, thereby contributing to broader discussions on physical and mental well-being. Similarly, Chen and Zhou (2022) employ an AI and intelligent robotics enabled human motion recognition algorithm, based on Hidden Markov Models (HMM), to analyze students' exercise behavior. Their findings inform the development of a structured model for college health education, thereby reinforcing the role of AI in promoting physical activity. The third representative article is by Deng *et al*. (2022) who integrate big data, Internet Plus, and AI to develop an intelligent governance system aimed at enhancing the physical health of college students. Their work highlights how advanced technologies can be systematically applied to strengthen health management frameworks within educational settings. Thus, within the theme of physical and mental health, these studies illustrate how AI and related technologies can be leveraged to establish institutional frameworks for fostering healthier lifestyle and improving overall student well-being.

The third extracted topic relates to social media sentiment analysis, with a representative example being the study by Galiano-Coronil *et al*. (2024). Using a combination of data mining, sentiment analysis, and content analysis, the authors examine social media communications of leading socially responsible companies in Spain. Their findings reveal that sports emerge as one of the most frequently addressed themes, underscoring the importance of sports-related discourse within corporate digital communication. The second study illustrative of Topic 3 is by Xie and Wang (2024), who employ artificial neural networks in combination with structural equation modeling to examine how AI can optimize the design and functionality of sports apps. Their findings highlight how improved app experiences can enhance user satisfaction and engagement, thereby reinforcing the role of AI in shaping digital interactions within the sports ecosystem. The third study reflective of this topic is Jain *et al*. (2017), who develop an AI-based system for emotion extraction from multilingual social media posts related to sports events. Using data from the 2015 Indian Premier League (IPL), the study illustrates how fan emotions and reactions expressed online can be harnessed through AI-driven techniques to determine their sentiments towards a sports event. Taken together, the above-mentioned studies underscore the growing relevance of AI-enabled sentiment analysis in understanding sports discourse on social media.

Representative studies for the fourth extracted topic, tactical tracking, emphasize the analysis of movement data, tactical behaviour, and decision-making in sports. Tuyls *et al*. (2021) provide an overarching perspective by demonstrating how statistics, game theory, and computer vision can be combined with AI to analyze soccer at a granular level, including passing networks, player positioning, and strategic decision-making. The second study exemplifying tactical tracking is by Goes *et al*. (2019). The authors work at the intersection of data science and sports science to quantify effectiveness of passes between players during soccer matches by using tracking data. Their approach provides valuable insight into tactical performance during a soccer match. Finally, Wu and Swartz (2023a) complete the trio of studies reflecting tactical tracking. Their research utilizes player tracking data derived from Cartesian coordinates to measure player speed, a key tactical performance indicator. By applying principles of exploratory data analysis, the study improves the reliability of speed estimation, thereby strengthening its contribution to sports analytics. In conjunction with each other, the aforementioned studies illustrate the contribution of tactical tracking research as a key constituent of AI-driven sports analytics.

**Table I.** Representative documents for each topic

| # | Topic | Representative study |
|---|-------|----------------------|
| 1 | Performance modelling | Maanijou, R., & Mirroshandel, S. A. (2019). Introducing an expert system for prediction of soccer player ranking using ensemble learning. *Neural Computing and Applications, 31*(12), 9157-9174. |
| | | Al-Asadi, M. A., & Tasdemır, S. (2022). Predict the value of football players using FIFA video game data and machine learning techniques. *IEEE access, 10*, 22631-22645. |
| | | Constantinou, A. C. (2019). Dolores: a model that predicts football match outcomes from all over the world. *Machine Learning, 108*(1), 49-75. |
| 2 | Physical and mental health | Peng, X., & Tang, L. (2021). Exploring the characteristics of physical exercise in students and the path of health education. *Frontiers in Psychology, 12*, 663922. |
| | | Chen, M., & Zhou, Y. (2022). Analysis of students' sports exercise behavior and health education strategy using visual perception–motion recognition algorithm. *Frontiers in Psychology, 13*, 829432. |
| | | Deng, C., Yu, Q., Luo, G., Zhao, Z., & Li, Y. (2022). Big data-driven intelligent governance of college students' physical health: System and strategy. *Frontiers in Public Health, 10*, 924025. |
| 3 | Social media sentiment analysis | Galiano-Coronil, A., Aguirre Montero, A., López Sánchez, J. A., & Díaz Ortega, R. (2024). Exploring social responsibility, social marketing and happiness using artificial intelligence, automated text analysis and correspondence analysis. *Management Decision, 62*(2), 549-574. |

| | | Xie, G., & Wang, X. (2024). Exploring the impact of AI enhancement on the sports app community: Analyzing human-computer interaction and social factors using a hybrid SEM-ANN approach. *International Journal of Human–Computer Interaction*, 1-22. |
|---|---|---|
| | | Jain, V. K., Kumar, S., & Fernandes, S. L. (2017). Extraction of emotions from multilingual text using intelligent text processing and computational linguistics. *Journal of Computational Science, 21*, 316-326. |
| 4 | Tactical tracking | Tuyls, K., Omidshafiei, S., Muller, P., Wang, Z., Connor, J., Hennes, D., ... & Hassabis, D. (2021). Game plan: What AI can do for football, and what football can do for AI. *Journal of Artificial Intelligence Research, 71*, 41-88. |
| | | Goes, F. R., Kempe, M., Meerhoff, L. A., & Lemmink, K. A. (2019). Not every pass can be an assist: a data-driven model to measure pass effectiveness in professional soccer matches. *Big Data, 7*(1), 57-70. |
| | | Wu, L. Y., & Swartz, T. B. (2023). The calculation of player speed from tracking data. *International Journal of Sports Science & Coaching, 18*(2), 516-522. |

Source: Author's own creation

*Topic mapping*

The inter topic distance map (Figure 5) presents interconnections among the extracted topics and highlights their relative significance. We employed BERTopic to generate a two-dimensional inter topic distance map that visualizes both the relative significance of the extracted topics and the semantic relationships among them. In this map, each topic is represented as a circle, with its size indicating relative significance and its spatial position reflecting semantic proximity or dissimilarity. Topics with similar word distributions are positioned closer together, denoting greater semantic similarity. As illustrated in the inter topic distance map, the relative significance of the topics is as follows: performance modelling, physical and mental health, social media sentiment analysis, and tactical tracking. The inter topic distance map demonstrates that the four extracted topics are largely distinct, as indicated by the spacing between their respective circles. The two most prominent areas, performance modelling and physical and mental health, are positioned relatively far apart from each other, underscoring their thematic divergence. At the same time, closer proximities are observed between physical and mental health and tactical tracking, as well as between performance modelling and social media sentiment analysis, suggesting areas of conceptual overlap. Overall, the map reveals that research on AI and analytics in sports clusters into four relatively distinct yet interrelated domains.

**Figure 5.** Inter topic distance map

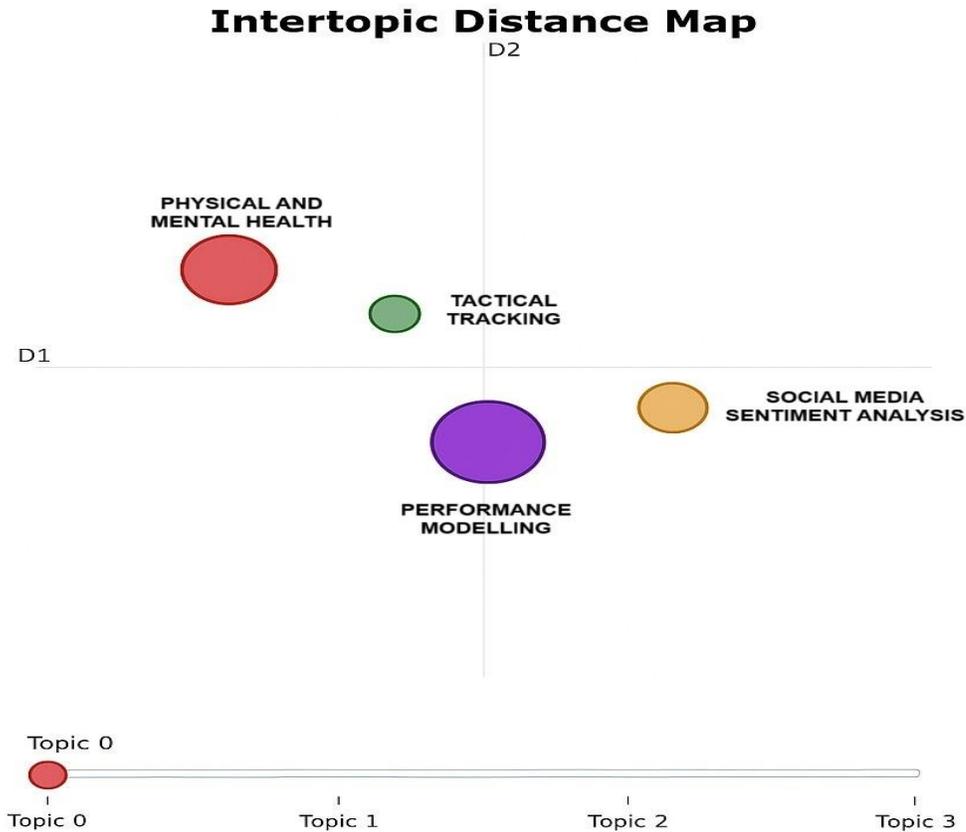

Source: Author's own creation

**Avenues for future research**

This study employed topic modelling algorithm of BERTopic to uncover latent themes in existing research on the application of AI and analytics in sports. The analysis provides insights into gaps within the current body of knowledge. Drawing on the inter topic distance map (Figure 5), which illustrates the relative importance and relationships of the four extracted topics, we propose the following avenues for future research. To begin with, the application of AI and analytics to tactical tracking in sports remains underexplored, as indicated by the relatively small size of its circle in the inter topic distance map. Tactical tracking has emerged as an increasingly important area of sports research, as it enables more effective tactical decision-making (Tuyls et al., 2021) and contributes to superior competitive performance (Ramos et al., 2024). Existing studies have applied AI and analytical techniques to examine a range of tracking data, including pass effectiveness (Goes et al., 2019), player positioning (Goes et al., 2021), speed (Wu & Swartz, 2023a), defensive velocity (Wu & Swartz, 2023b), pace of play (Shen et al., 2022), and effective playing space (Ötting & Karlis, 2023). While these investigations illustrate the breadth of tactical insights that can be derived, opportunities exist in following new lines of inquiry. First, future research should prioritize the integration of multimodal tracking data into comprehensive analytical frameworks. Such approaches would generate richer and more nuanced insights into tactical performance and competitiveness. Second, with advancements in tracking technologies, the development of real-time AI-based tactical decision support systems warrants exploration. These systems could assist coaches in adjusting tactical strategies during matches, thereby enhancing competitive advantage. Finally, tactical tracking research should be extended beyond soccer to other sports such as cricket, field hockey, and volleyball. Broadening the scope in this way would not only uncover sport-specific dynamics but also contribute to the development of more generalizable AI models. Exploring these directions offers considerable potential for advancing the application of AI and analytics in tactical tracking.

      Secondly, analysis of sports-related sentiment on social media using AI and analytics has emerged as an important research theme in extant research. Beyond tactics and game plans, sport holds significant experiential value owing to the emotions it generates (Qian *et al*., 2024). Viewer experience (Liu *et al*., 2024) and social interaction (Depper & Howe, 2017) are integral dimensions

of this emotional landscape. Both players (Bigsby *et al*., 2019) and fans (Chang, 2019; Deep Prakash & Majumdar, 2023) actively engage in emotional expression on social media platforms, creating a rich corpus of data for analysis. These online expressions provide a valuable resource for applying AI and analytics to better understand fan emotions and user experience (Jain *et al*., 2017; Xie & Wang, 2024). Despite advances in this area, further research is required to pursue a more nuanced exploration of sentiment dynamics in the sports domain. Possible research inquiries include the following. First, although sport is a global phenomenon, cultural context plays a central role in shaping sporting traditions and sentiment expressions (Deep Prakash & Majumdar, 2023). Future research should therefore move beyond sentiments expressed in English to capture multilingual and cross-cultural perspectives. Second, to better reflect the diversity of fan emotions, sentiment analysis should extend beyond textual data to incorporate images, videos, emojis, gifs, and live streams. Third, future work should explore the relationship between fan sentiment and behavioral or market outcomes, which may manifest positively, through loyalty and revenue generation, or negatively, in the form of aggression, physical harm, or property damage. Addressing these directions would not only advance methodological sophistication but also enhance the practical significance of sentiment analysis in the sports domain.

The third extracted topic concerns research on physical and mental health in sports enabled by AI and analytics. While investigations into physical health outcomes have long been central to sports research (Deng *et al*., 2022; Peng & Tang, 2021), growing recognition of mental health challenges among athletes has brought this dimension to the forefront in recent years (Liang *et al*., 2022). Existing studies have examined areas such as injury prediction using wearable data (Zadeh *et al*., 2021), the influence of athletic intelligence quotient on performance (Bowman *et al*., 2021), and the neurological effects of repetitive subconcussive head impacts (Miller *et al*., 2021).

Nevertheless, further inquiry is needed given the critical interdependence between physical and mental health and their combined impact on athletic performance (Glandorf *et al*., 2025). Against this backdrop, several avenues for future research can be identified. First, AI and analytics should be applied towards modelling preventive detection of physical injuries by drawing upon biomechanical or physiological indicators that precede critical incidents. Second, there is significant potential to use AI for modelling the risk of mental health challenges through multimodal data sources such as social media activity, sleep patterns, and even conversational cues. Such research would enable timely interventions to mitigate stress, burnout, and depression among athletes. Third, while most existing studies treat physical and mental health separately, future work should develop integrated AI frameworks that capture the interplay between physical and emotional well-being. An exploration into the aforementioned research directions would further contribute towards utilization of AI and analytics to safeguard sportsperson well-being.

The fourth extracted topic, performance modelling, represents the area where the application of AI and analytics in sports has received the most scholarly attention. Existing studies have developed models to predict player rankings (Maanijou & Mirroshandel, 2019), estimate player market values (Al-Asadi & Tasdemir, 2022), and forecast match outcomes (Constantinou, 2019). Performance prediction has been explored across a variety of contexts, including baseball (Elitzur, 2020), soccer (Hubáček et al., 2019), ice hockey (Gu et al., 2019), American football (Hsu, 2020), and more recently, e-sports (Minami et al., 2024). While this body of work demonstrates the versatility of AI-driven performance modelling, further inquiry is warranted to examine how such approaches can further add value to the sports administrators' decision-making process. Against this backdrop, the following research avenues are proposed. First, future research should endeavour to build AI and analytics enabled performance models that could holistically capture technical, tactical, physical, and psychological facets at individual and team levels. Such a comprehensive dashboard would provide enriched understanding of player and team capabilities. Second, expanding upon the previous research direction, performance modelling framework could be further enhanced by incorporating situational factors such as, game context, opponent quality, and environmental conditions. Third, future research should focus on connecting performance models with actionable decision support systems for productive backward integration in terms of talent identification, development, and recruitment. Treading on the aforesaid research directions would expand the domain beyond the current horizons.

**Discussion**

Leveraging the advanced topic modelling algorithm of BERTopic, the study thus extracted four underlying topics from extant research on utilization of AI and analytics in sports: performance modelling, physical and mental health, social media sentiment analysis, and tactical tracking. While extraction of latent topics is novel, the following discussion elaborates upon their logical alignment with fundamental principles of sports research.

*Performance modelling*

Prior research underscores a strong association between AI, analytics, and performance modelling in sports (Al-Asadi & Tasdemır, 2022; Maanijou & Mirroshandel, 2019). Performance represents a foundational principle of sports research, as it directly influences competitive outcomes, strategic decision-making, and resource allocation. With the advent of AI-powered algorithms and advanced analytical techniques, modelling performance outcomes emerges as a logical extension of this principle. Scholars have applied AI and analytics to predict a wide range of performance-related phenomena, including match outcomes (Hsu, 2020), event attendance (Nguyen *et al*., 2022), Olympic medal distribution (Schlembach *et al*., 2022), gifting behaviour in live sports streaming (Liu *et al*., 2022), and player market value (Inan & Cavas, 2021). As big data becomes increasingly integral to sports analytics (Swartz, 2020), harnessing the power of AI and advanced analytics (Sakthivel *et al*., 2024) to generate actionable insights for performance modelling is not only expected but essential for advancing both scholarly inquiry and practice (Herold *et al*., 2024).

*Physical and mental health*

The physical and mental well-being of athletes is a critical principle of sports research, as it directly influences their capacity to train, compete, and sustain performance. Traditionally, greater attention has been directed towards physical health concerns such as physiological monitoring, injury prevention, and rehabilitation (Chen & Zhou, 2022; Peng & Tang, 2021; Deng *et al*., 2022). More recently, however, growing recognition of mental health challenges, including stress, burnout, and depression, has expanded the scope of inquiry (Liang *et al*., 2022; Sarlis & Tjortjis, 2024). Advances in AI and analytics have significantly enhanced the ability of decision-makers and caregivers to prevent, detect, and treat these issues. For instance, IoT-enabled fitness devices have been used to track health metrics (Farrokhi *et al*., 2021), deliver real-time feedback (Dorschky *et al*., 2023), and forecast injuries (Qi *et al*., 2024; Zadeh *et al*., 2021). Similarly, AI models have been applied to examine the role of athletic intelligence quotient in performance (Bowman *et al*., 2021) and to assess the neurological consequences of repetitive subconcussive impacts (Miller *et al*., 2021). Collectively, these developments suggest that integration of AI and analytics with big data from athlete monitoring represents a logical extension of the principle that safeguarding both physical and psychological health is foundational to sustained performance (Glandorf *et al*., 2025).

*Social media sentiment analysis*

Social media platforms provide a pervasive avenue for individuals to voice opinions through likes, comments, and shares, and to promote themselves or others via posts, pictures, and videos (Roggeveen & Grewal, 2016). Their central role in shaping customer engagement and experience is well recognized (Lim & Rasul, 2022; Mishra, 2025), largely due to their capacity to create participatory and immersive experiences. Sports and social media are closely intertwined, as both are hedonic and community-driven activities thriving on shared emotions and fun (Oh & Pham, 2022). It is therefore expected that fans, players, and stakeholders are likely to express sentiments on social media, influencing outcomes such as game attendance (Gong *et al*., 2021), recruitment decisions in college American football (Bigsby *et al*., 2019), and broader patterns of fan engagement (Deep Prakash & Majumdar, 2023). AI and analytics have further strengthened this linkage by enabling systematic extraction and analysis of emotional expressions, offering profound insights into user experiences and interactions (Jain *et al*., 2017; Xie & Wang, 2024). Consequently, AI and analytics driven sentiment analysis has become a central stream of research, advancing our understanding of emotional engagement and fan experiences within the sports domain.

*Tactical tracking*

The integration of AI, analytics, and big data has enabled sports researchers and practitioners to analyze performance at a granular level, capturing individual player actions across time and context. This shift has advanced tactical tracking from a purely observational exercise to a systematic, data-driven approach aligned with defined performance objectives. Goes *et al*. (2019) note that nearly all professional soccer teams now employ tracking technologies to monitor players during both matches and training. Applications of these systems include generating strategic insights into player speed and acceleration (Wu & Swartz, 2023a), analyzing penalty kick strategies (Tuyls *et al*., 2021), evaluating passing effectiveness (Goes *et al*., 2019), and identifying optimal actions in different game situations (Rahimian & Toka, 2024). Such data allows analysts to move beyond headline events like goals to uncover the micro-dynamics of play. Accordingly, the growing body of research on AI-enabled tactical tracking reflects the enduring principle of sports research: leveraging systematic evidence to inform tactical behavior, decision-making, and competitive advantage.

**Conclusion**

This study adopts a methodological blend of SLR and topic modelling to provide a holistic assessment of how AI and analytics have been utilized in the sports domain (Figure 1). Guided by the PRISMA protocol (Moher et al., 2009), we identified 204 relevant articles published between 2002 and 2024. The sample, validated through a word cloud of dominant terms such as "sport," "model," "data," "player," "machine learning," and "performance," reflects its representativeness of AI and analytics focused research within sports domain (Figure 2). A descriptive overview further reveals that leading contributions originate from journals such as *Frontiers in Psychology*, *International Journal of Sports Science & Coaching*, *International Journal of Forecasting*, *Sports Management Review*, *Big Data*, *PLOS One*, *Applied Sciences-Basel*, and *Sustainability* underscoring the multidisciplinary reach of this field (Figure 3). To extend the review, we applied BERTopic (Grootendorst, 2022; Mishra, 2025) to abstracts of the 204 articles, which revealed four distinct themes: performance modelling, physical and mental health, social media sentiment analysis, and tactical tracking. These themes, ordered by relative importance as visualized in the inter topic distance map (Figure 5), were further characterized through their associated terms (Figure 4) and illustrated with representative studies (Table 1). In conclusion, we believe that the insights and directions offered in this study would advance scholarly inquiry and foster a more holistic understanding of how AI and analytics can transform the domain of sports.

**References**

Al-Asadi, M.A. and Tasdemir, S. (2022), "Predict the value of football players using FIFA video game data and machine learning techniques", *IEEE Access*, Vol. 10, pp. 22631–22645.

Asmussen, C.B. and Møller, C. (2019), "Smart literature review: a practical topic modelling approach to exploratory literature review", *Journal of Big Data*, Vol. 6 No. 1, pp. 1–18.

Bag, S., Dhamija, P., Singh, R.K., Rahman, M.S. and Sreedharan, V.R. (2023), "Big data analytics and artificial intelligence technologies-based collaborative platform empowering absorptive capacity in health care supply chain: an empirical study", *Journal of Business Research*, Vol. 154, p. 113315.

Basole, R.C., Park, H. and Seuss, C.D. (2024), "Complex business ecosystem intelligence using AI-powered visual analytics", *Decision Support Systems*, Vol. 178, p. 114133.

Bigsby, K.G., Ohlmann, J.W. and Zhao, K. (2019), "Keeping it 100: social media and self-presentation in college football recruiting", *Big Data*, Vol. 7 No. 1, pp. 3–20.

Blei, D.M. and Lafferty, J.D. (2009), "Topic models", in Srivastava, A.N. and Sahami, M. (Eds), *Text Mining: Classification, Clustering, and Applications*, Taylor and Francis, London, pp. 71–94.

Blei, D.M., Ng, A.Y. and Jordan, M.I. (2003), "Latent Dirichlet allocation", *Journal of Machine Learning Research*, Vol. 3, pp. 993–1022.

Borgman, C.L. and Furner, J. (2002), "Scholarly communication and bibliometrics", *Annual Review of Information Science and Technology*, Vol. 36 No. 1, pp. 2–72.

Bowman, J.K., Boone, R.T., Goldman, S. and Auerbach, A. (2021), "The athletic intelligence quotient and performance outcomes in professional baseball", *Frontiers in Psychology*, Vol. 12, p. 629827.

Bradlow, E.T., Gangwar, M., Kopalle, P. and Voleti, S. (2017), "The role of big data and predictive analytics in retailing", *Journal of Retailing*, Vol. 93 No. 1, pp. 79–95.

Broby, D. (2022), "The use of predictive analytics in finance", *The Journal of Finance and Data Science*, Vol. 8, pp. 145–161.

Casals, M. and Daunis-i-Estadella, P. (2023), "Violinboxplot and enhanced radar plot as components of effective graphical dashboards: an educational example of sports analytics", *International Journal of Sports Science & Coaching*, Vol. 18 No. 2, pp. 572–583.

Chang, Y. (2019), "Spectators' emotional responses in tweets during the Super Bowl 50 game", *Sport Management Review*, Vol. 22 No. 3, pp. 348–362.

Chen, M. and Zhou, Y. (2022), "Analysis of students' sports exercise behavior and health education strategy using visual perception–motion recognition algorithm", *Frontiers in Psychology*, Vol. 13, p. 829432.

Chen, X., Zou, D., Xie, H., Cheng, G. and Liu, C. (2022), "Two decades of artificial intelligence in education", *Educational Technology & Society*, Vol. 25 No. 1, pp. 28–47.

Chmait, N. and Westerbeek, H. (2021), "Artificial intelligence and machine learning in sport research: an introduction for non-data scientists", *Frontiers in Sports and Active Living*, Vol. 3, p. 682287.

Constantinou, A.C. (2019), "Dolores: a model that predicts football match outcomes from all over the world", *Machine Learning*, Vol. 108 No. 1, pp. 49–75.

Crain, S.P., Zhou, K., Yang, S.H. and Zha, H. (2012), "Dimensionality reduction and topic modeling: from latent semantic indexing to latent Dirichlet allocation and beyond", in Aggarwal, C. and Zhai, C. (Eds), *Mining Text Data*, Springer, Boston, MA.

Davenport, T.H. and Harris, J.G. (2007), *Competing on Analytics: The New Science of Winning*, Harvard Business School Press.

Deep Prakash, C. and Majumdar, A. (2023), "Predicting sports fans' engagement with culturally aligned social media content: a language expectancy perspective", *Journal of Retailing and Consumer Services*, Vol. 75, p. 103457.

Deng, C., Yu, Q., Luo, G., Zhao, Z. and Li, Y. (2022), "Big data-driven intelligent governance of college students' physical health: system and strategy", Frontiers in Public Health, Vol. 10, p. 924025.

Depper, A. and Howe, P.D. (2017), "Are we fit yet? English adolescent girls' experiences of health and fitness apps", Health Sociology Review, Vol. 26 No. 1, pp. 98–112.

DiMaggio, P., Nag, M. and Blei, D. (2013), "Exploiting affinities between topic modeling and the sociological perspective on culture: application to newspaper coverage of US government arts funding", Poetics, Vol. 41 No. 6, pp. 570–606.

Dindorf, C., Bartaguiz, E., Gassmann, F. and Fröhlich, M. (2022), "Conceptual structure and current trends in artificial intelligence, machine learning, and deep learning research in sports: a bibliometric review", International Journal of Environmental Research and Public Health, Vol. 20 No. 1, p. 173.

Elitzur, R. (2020), "Data analytics effects in major league baseball", Omega, Vol. 90, p. 102001.

Erevelles, S., Fukawa, N. and Swayne, L. (2016), "Big data consumer analytics and the transformation of marketing", Journal of Business Research, Vol. 69 No. 2, pp. 897–904.

Farrokhi, A., Farahbakhsh, R., Rezazadeh, J. and Minerva, R. (2021), "Application of Internet of Things and artificial intelligence for smart fitness: a survey", Computer Networks, Vol. 189, p. 107859.

Galiano-Coronil, A., Aguirre Montero, A., López Sánchez, J.A. and Díaz Ortega, R. (2024), "Exploring social responsibility, social marketing and happiness using artificial intelligence, automated text analysis and correspondence analysis", Management Decision, Vol. 62 No. 2, pp. 549–574.

Germann, F., Lilien, G.L. and Rangaswamy, A. (2013), "Performance implications of deploying marketing analytics", International Journal of Research in Marketing, Vol. 30 No. 2, pp. 114–128.

Ghosh, I., Ramasamy Ramamurthy, S., Chakma, A. and Roy, N. (2023), "Sports analytics review: artificial intelligence applications, emerging technologies, and algorithmic perspective", Wiley Interdisciplinary Reviews: Data Mining and Knowledge Discovery, Vol. 13 No. 5, p. e1496.

Glandorf, H.L., Madigan, D.J., Kavanagh, O. and Mallinson-Howard, S.H. (2025), "Mental and physical health outcomes of burnout in athletes: a systematic review and meta-analysis", International Review of Sport and Exercise Psychology, Vol. 18 No. 1, pp. 372–416.

Goes, F.R., Kempe, M., Meerhoff, L.A. and Lemmink, K.A. (2019), "Not every pass can be an assist: a data-driven model to measure pass effectiveness in professional soccer matches", Big Data, Vol. 7 No. 1, pp. 57–70.

Goes, F.R., Meerhoff, L.A., Bueno, M.J.O., Rodrigues, D.M., Moura, F.A., Brink, M.S. and Lemmink, K.A.P.M. (2021), "Unlocking the potential of big data to support tactical performance analysis in professional soccer: a systematic review", European Journal of Sport Science, Vol. 21 No. 4, pp. 481–496.

Gong, H., Watanabe, N.M., Soebbing, B.P., Brown, M.T. and Nagel, M.S. (2021), "Do consumer perceptions of tanking impact attendance at National Basketball Association games? A sentiment analysis approach", Journal of Sport Management, Vol. 35 No. 3, pp. 254–265.

Grootendorst, M. (2022), "BERTopic: neural topic modeling with a class-based TF-IDF procedure", arXiv preprint, arXiv:2203.05794.

Gu, W., Foster, K., Shang, J. and Wei, L. (2019), "A game-predicting expert system using big data and machine learning", Expert Systems with Applications, Vol. 130, pp. 293–305.

Guo, Y., Barnes, S.J. and Jia, Q. (2017), "Mining meaning from online ratings and reviews: tourist satisfaction analysis using latent Dirichlet allocation", Tourism Management, Vol. 59, pp. 467–483.

Hannigan, T.R., Haans, R.F., Vakili, K., Tchalian, H., Glaser, V.L., Wang, M.S. and Jennings, P.D. (2019), "Topic modeling in management research: rendering new theory from textual data", Academy of Management Annals, Vol. 13 No. 2, pp. 586–632.

Herold, E., Singh, A., Feodoroff, B. and Breuer, C. (2024), "Data-driven message optimization in dynamic sports media: an artificial intelligence approach to predict consumer response", Sport Management Review, Vol. 27 No. 5, pp. 793–816.

Hartmann, N.N., Wieland, H., Gustafson, B. and Habel, J. (2024), "Research on sales and ethics: mapping the past and charting the future", Journal of the Academy of Marketing Science, Vol. 52 No. 3, pp. 653–671.

Hiebl, M.R. (2021), "Sample selection in systematic literature reviews of management research", Organizational Research Methods, Vol. 26 No. 2, pp. 229–261.

Hsu, Y.C. (2020), "Using machine learning and candlestick patterns to predict the outcomes of American football games", Applied Sciences, Vol. 10 No. 13, p. 4484.

Hubáček, O., Šourek, G. and Železný, F. (2019), "Learning to predict soccer results from relational data with gradient boosted trees", Machine Learning, Vol. 108 No. 1, pp. 29–47.

Hulland, J. and Houston, M.B. (2020), "Why systematic review papers and meta-analyses matter: an introduction to the special issue on generalizations in marketing", Journal of the Academy of Marketing Science, Vol. 48, pp. 351–359.

Inan, T. and Cavas, L. (2021), "Estimation of market values of football players through artificial neural network: a model study from the Turkish Super League", Applied Artificial Intelligence, Vol. 35 No. 13, pp. 1022–1042.

Jacobi, C., van Atteveldt, W. and Welbers, K. (2016), "Quantitative analysis of large amounts of journalistic texts using topic modelling", Digital Journalism, Vol. 4 No. 1, pp. 89–106.

Jacsó, P. (2005), "Google Scholar: the pros and the cons", Online Information Review, Vol. 29 No. 2, pp. 208–214.

Jain, V.K., Kumar, S. and Fernandes, S.L. (2017), "Extraction of emotions from multilingual text using intelligent text processing and computational linguistics", Journal of Computational Science, Vol. 21, pp. 316–326.

Klaus, P. and Kuppelwieser, V. (2021), "Guiding directions and propositions: placing dynamics at the heart of customer experience (CX) research", Journal of Retailing and Consumer Services, Vol. 59, p. 102429.

Liang, L., Zheng, Y., Ge, Q. and Zhang, F. (2022), "Exploration and strategy analysis of mental health education for students in sports majors in the era of artificial intelligence", Frontiers in Psychology, Vol. 12, p. 762725.

Lim, W.M. and Rasul, T. (2022), "Customer engagement and social media: revisiting the past to inform the future", Journal of Business Research, Vol. 148, pp. 325–342.

Liu, H., Tan, K.H. and Pawar, K. (2022), "Predicting viewer gifting behavior in sports live streaming platforms: the impact of viewer perception and satisfaction", Journal of Business Research, Vol. 144, pp. 599–613.

Liu, H., Chung, L., Tan, K.H. and Peng, B. (2024), "I want to view it my way! How viewer engagement shapes the value co-creation on sports live streaming platform", Journal of Business Research, Vol. 170, p. 114331.

Maanijou, R. and Mirroshandel, S.A. (2019), "Introducing an expert system for prediction of soccer player ranking using ensemble learning", Neural Computing and Applications, Vol. 31 No. 12, pp. 9157–9174.

Miller, L.E., Urban, J.E., Davenport, E.M., Powers, A.K., Whitlow, C.T., Maldjian, J.A. and Stitzel, J.D. (2021), "Brain strain: computational model-based metrics for head impact exposure and injury correlation", Annals of Biomedical Engineering, Vol. 49 No. 3, pp. 1083–1096.

Minami, S., Koyama, H., Watanabe, K., Saijo, N. and Kashino, M. (2024), "Prediction of esports competition outcomes using EEG data from expert players", Computers in Human Behavior, Vol. 160, p. 108351.

Mishra, M. (2022), "Customer experience: extracting topics from tweets", International Journal of Market Research, Vol. 64 No. 3, pp. 334–353.

Mishra, M. (2025), "A holistic review of customer experience research: topic modelling using BERTopic", Marketing Intelligence & Planning, Vol. 43 No. 4, pp. 802–820.

Mishra, M. and Mohapatra, S. (2025), "Exploring extant consumer research on narcissism: a topic modelling approach", Journal of Consumer Marketing, Vol. 42 No. 6, pp. 861–873.

Mishra, M. and Mund, P. (2024), "Fifty-two years of consumer research based on social exchange theory: a review and research agenda using topic modeling", International Journal of Consumer Studies, Vol. 48 No. 4, p. e13074.

Moher, D., Liberati, A., Tetzlaff, J. and Altman, D.G. (2009), "Preferred reporting items for systematic reviews and meta-analyses: the PRISMA statement", *British Medical Journal*, Vol. 339(b2535), pp. 332–336.

Nguyen, J.K., Karg, A., Valadkhani, A. and McDonald, H. (2022), "Predicting individual event attendance with machine learning: a 'step-forward' approach", *Applied Economics*, Vol. 54 No. 27, pp. 3138–3153.

Oesper, L., Merico, D., Isserlin, R. and Bader, G.D. (2011), "WordCloud: a Cytoscape plugin to create a visual semantic summary of networks", *Source Code for Biology and Medicine*, Vol. 6 No. 1, p. 7.

Oh, T.T. and Pham, M.T. (2022), "A liberating-engagement theory of consumer fun", *Journal of Consumer Research*, Vol. 49 No. 1, pp. 46–73.

Ötting, M. and Karlis, D. (2023), "Football tracking data: a copula-based hidden Markov model for classification of tactics in football", *Annals of Operations Research*, Vol. 325, pp. 167–183.

Palmatier, R.W., Houston, M.B. and Hulland, J. (2018), "Review articles: purpose, process, and structure", *Journal of the Academy of Marketing Science*, Vol. 46 No. 1, pp. 1–15.

Paul, J. and Criado, A.R. (2020), "The art of writing literature review: what do we know and what do we need to know?", *International Business Review*, Vol. 29 No. 4, p. 101717.

Paul, J., Lim, W.M., O'Cass, A., Hao, A.W. and Bresciani, S. (2021), "Scientific procedures and rationales for systematic literature reviews (SPAR-4-SLR)", *International Journal of Consumer Studies*, Vol. 45 No. 4, pp. 1–16.

Peng, X. and Tang, L. (2021), "Exploring the characteristics of physical exercise in students and the path of health education", *Frontiers in Psychology*, Vol. 12, p. 663922.

Qi, Y., Sajadi, S.M., Baghaei, S., Rezaei, R. and Li, W. (2024), "Digital technologies in sports: opportunities, challenges, and strategies for safeguarding athlete wellbeing and competitive integrity in the digital era", *Technology in Society*, Vol. 77, p. 102496.

Qian, T.Y., Li, W., Gong, H., Seifried, C. and Xu, C. (2024), "Experience is all you need: a large language model application of fine-tuned GPT-3.5 and RoBERTa for aspect-based sentiment analysis of college football stadium reviews", *Sport Management Review*, Vol. 28 No. 1, pp. 1–25.

Raddats, C., Kowalkowski, C., Benedettini, O., Burton, J. and Gebauer, H. (2019), "Servitization: a contemporary thematic review of four major research streams", *Industrial Marketing Management*, Vol. 83, pp. 207–223.

Rahimian, P. and Toka, L. (2024), "A data-driven approach to assist offensive and defensive players in optimal decision making", *International Journal of Sports Science & Coaching*, Vol. 19 No. 1, pp. 245–256.

Rajšp, A. and Fister Jr, I. (2020), "A systematic literature review of intelligent data analysis methods for smart sport training", *Applied Sciences*, Vol. 10 No. 9, p. 3013.

Ramos, A., Davids, K., Coutinho, P. and Mesquita, I. (2024), "Tracking relations between development of tactical knowledge and tactical behaviour: a season-long action research study", *Physical Education and Sport Pedagogy*, Vol. 29 No. 4, pp. 346–360.

Roggeveen, A.L. and Grewal, D. (2016), "Engaging customers: the wheel of social media engagement", *Journal of Consumer Marketing*, Vol. 33 No. 2.

Sakthivel, V., Pravakar, D. and Prakash, P. (2024), "Harnessing the power of artificial intelligence and data science", in *Advancement of Data Processing Methods for Artifical and Computing Intelligence*, River Publishers, pp. 305–327.

Sarlis, V. and Tjortjis, C. (2024), "Sports analytics: data mining to uncover NBA player position, age, and injury impact on performance and economics", *Information*, Vol. 15 No. 4, p. 242.

Sarlis, V. and Tjortjis, C. (2020), "Sports analytics—evaluation of basketball players and team performance", *Information Systems*, Vol. 93, p. 101562.

Schlembach, C., Schmidt, S.L., Schreyer, D. and Wunderlich, L. (2022), "Forecasting the Olympic medal distribution – a socioeconomic machine learning model", *Technological Forecasting and Social Change*, Vol. 175, p. 121314.

Seyedan, M. and Mafakheri, F. (2020), "Predictive big data analytics for supply chain demand forecasting: methods, applications, and research opportunities", *Journal of Big Data*, Vol. 7 No. 1, p. 53.

Shen, E., Santo, S. and Akande, O. (2022), "Analysing pace-of-play in soccer using spatio-temporal event data", *Journal of Sports Analytics*, Vol. 8 No. 2, pp. 127–139.

Sjödin, D., Parida, V. and Kohtamäki, M. (2023), "Artificial intelligence enabling circular business model innovation in digital servitization: conceptualizing dynamic capabilities, AI capacities, business models and effects", *Technological Forecasting and Social Change*, Vol. 197, p. 122903.

Snyder, H. (2019), "Literature review as a research methodology: an overview and guidelines", *Journal of Business Research*, Vol. 104, pp. 333–339.

Spanaki, K., Karafili, E. and Despoudi, S. (2021), "AI applications of data sharing in agriculture 4.0: a framework for role-based data access control", *International Journal of Information Management*, Vol. 59, p. 102350.

Stead, S., Wetzels, R., Wetzels, M., Odekerken-Schröder, G. and Mahr, D. (2022), "Toward multisensory customer experiences: a cross-disciplinary bibliometric review and future research directions", *Journal of Service Research*, Vol. 25 No. 3, pp. 440–459.

Stryker, S. and Kavlakoglu, E. (2025), "What is artificial intelligence (AI)?", *IBM Think Blog*, accessed 20 June 2025.

Swartz, T.B. (2020), "Where should I publish my sports paper?", *The American Statistician*, Vol. 74 No. 2, pp. 103–108.

Tirunillai, S. and Tellis, G.J. (2014), "Mining marketing meaning from online chatter: strategic brand analysis of big data using latent Dirichlet allocation", *Journal of Marketing Research*, Vol. 51 No. 4, pp. 463–479.

Tranfield, D., Denyer, D. and Smart, P. (2003), "Towards a methodology for developing evidence-informed management knowledge by means of systematic review", *British Journal of Management*, Vol. 14 No. 3, pp. 207–222.

Tuyls, K., Omidshafiei, S., Muller, P., Wang, Z., Connor, J., Hennes, D. and Hassabis, D. (2021), "Game plan: what AI can do for football, and what football can do for AI", *Journal of Artificial Intelligence Research*, Vol. 71, pp. 41–88.
31

Vanhala, M., Lu, C., Peltonen, J., Sundqvist, S., Nummenmaa, J. and Järvelin, K. (2020), "The usage of large data sets in online consumer behaviour: a bibliometric and computational text-mining–driven analysis of previous research", *Journal of Business Research*, Vol. 106, pp. 46–59.
Watanabe, N.M., Shapiro, S. and Drayer, J. (2021), "Big data and analytics in sport management", *Journal of Sport Management*, Vol. 35 No. 3, pp. 197–202.
Wu, L.Y. and Swartz, T.B. (2023a), "The calculation of player speed from tracking data", *International Journal of Sports Science & Coaching*, Vol. 18 No. 2, pp. 516–522.
Wu, Y. and Swartz, T. (2023b), "Evaluation of off-the-ball actions in soccer", *Statistica Applicata - Italian Journal of Applied Statistics*, Vol. 35 No. 2.
Xie, G. and Wang, X. (2024), "Exploring the impact of AI enhancement on the sports app community: analysing human-computer interaction and social factors using a hybrid SEM-ANN approach", *International Journal of Human–Computer Interaction*, Vol. 41 No. 14, pp. 8734–8755.
Zadeh, A., Taylor, D., Bertsos, M., Tillman, T., Nosoudi, N. and Bruce, S. (2021), "Predicting sports injuries with wearable technology and data analysis", *Information Systems Frontiers*, Vol. 23 No. 4, pp. 1023–1037.
Zhao, J., Mao, J. and Tan, J. (2022), "Global trends and hotspots in research on extended reality in sports: a bibliometric analysis from 2000 to 2021", *Digital Health*, Vol. 8, p. 20552076221131141.